\documentclass[amsmath,amssymb,amssymb]{revtex4-1}

\usepackage{graphicx,epsfig,epstopdf}% Include figure files
\usepackage{mathrsfs}
\usepackage{dcolumn}% Align table columns on decimal point
\usepackage{bm}% bold math
\usepackage{natbib}
\begin{document}
\title{Cascading and Local-Field Effects in Non-Linear Optics Revisited; A Quantum-Field Picture Based on Exchange of Photons}
\author{Kochise Bennett}
\email{kcbennet@uci.edu}
\affiliation{Chemistry Department, University of California, Irvine, California 92697-2025, USA}
\author{Shaul Mukamel}
\email{smukamel@uci.edu}
\affiliation{Chemistry Department, University of California, Irvine, California 92697-2025, USA}
\date{\today}
\begin{abstract}
The semi-classical theory of radiation-matter coupling misses local-field effects that may alter the pulse time-ordering and cascading that leads to the generation of new signals.  These are then introduced macroscopically by solving Maxwell's equations. This procedure is convenient and intuitive but ad hoc.  We show that both effects emerge naturally by including coupling to quantum modes of the radiation field in the vacuum state to second order. This approach is systematic and suggests a more general class of corrections that only arise in a QED framework.  In the semi-classical theory, which only includes classical field modes, the susceptibility of a collection of $N$ non-interacting molecules is additive and scales as $N$.  Second-order coupling to a vacuum mode generates an effective retarded interaction that leads to cascading and local field effects both of which scale as $N^2$.
\end{abstract}
\maketitle
\section{introduction}
Spectroscopy seeks to use the optical response of matter to determine properties of the constituent molecules.  Since this typically involves probing a sample composed of many molecules, it is necessary to relate the response of the entire sample to that of a single molecule.  When the sample is sufficiently dilute, the signals from each molecule simply add and the total response is proportional to the molecular response \cite{mukbook, boydnonlinear}.  In denser samples, this picture must be corrected.
\par
The semi-classical approach involves a patch-up of macroscopic and microscopic levels of theory.  The direct expansion of the polarization for a system of non-interacting molecules is linear in $N$, as in a dilute sample; we call this the microscopic semi-classical expansion.  Higher-order effects are then included macroscopically in an ad hoc fashion by solving Maxwell's equations, resulting in cascading, which scales as $N^2$, and local-field corrections which scale as $N^2$ and higher (the Clausius-Mossotti relation) \cite{mukbook, scully1997quantum}; we call this corrected picture the macroscopic semi-classical expansion.  This phenomenological approach is very convenient and intuitive but is not systematic and one cannot be sure what effects are left unaccounted for.  Historically, the above effects were first discovered experimentally and then added to the formalism.  Local-field effects were included in off-resonant frequency domain susceptibilities to resolve discrepancies of the calculated absolute magnitude with experiment \cite{louie1975local, BoydLF, smith1997cancellation, Hache:86}.  In the time domain, they were required when unexpected signals showed that the time ordering of short pulses is scrambled by molecules with long-lived polarization (slow dephasing) \cite{DantusCascaded, cundiff2002time}.  Various other experimental observables are altered by local-field effects, such as the transmission/reflection of a thin film \cite{benedict1991reflection} and the Rabi oscillations of a quantum dot \cite{paspalakis2006local},  have been identified.  Similarly, cascading was introduced to account for new signals not included by the microscopic semi-classical approach \cite{mehlenbacher2009, TokmakoffFleming, kubarych2003}. Cascading signals arise when a molecule in the material interacts with the electromagnetic (EM) field and a polarization is produced which propagates to interact with another molecule from which the signal is ultimately detected. The effective response function for cascading signals therefore comes as a product of two lower-order response functions corresponding to the two molecules (i.e. $\chi^{(3)}$ like behavior can arise from a product of two factors of $\chi^{(2)}$, $\chi^{(5)}$ from a product of two factors of $\chi^{(3)}$, etc.). Cascading signals have the same wavevector and dependence on the incoming field amplitudes as the original signals, making them hard to distinguish. Various methods for separating out cascading signals from the higher-order process have been pursued \cite{golonzka2000, blank1999, gelin2013simple, zhao2011}. In the macroscopic semi-classical approach, this is calculated by creating a polarization and propagating with Maxwell's equations \cite{grimberg2002ultrafast, boydnonlinear}.
\par
In previous work, stimulated emission signals were analyzed from a microscopic perspective of a quantum radiation field and a number of expressions that were formerly obtained semi-classically were developed in a simpler manner \cite{roslyak2009, marx2008, schweigert2008} (more general introductions to the quantum nature of the radiation field can be found in \cite{cohen1992atom, milonni1994quantum, craig1998molecular, salam}). In this paper, we extend this formalism to vacuum-mediated interactions (VMI), and show how cascading and local-field processes, are caused by second-order in interactions with quantum modes.  In the emerging picture, one of a pair of molecules interacts with one or more EM fields before emitting into a vacuum mode.  The second molecule subsequently interacts with this vacuum mode and possibly other EM fields before producing a signal. Thus, while the sample is taken to be non-interacting, an effective interaction is nonetheless mediated via the vacuum field. This quantum-field approach introduces cascading and local-field corrections in an elegant way.  Everything is systematically related to the expansion order and no ambiguity remains regarding what processes have and have not been accounted for.  We find that
(\textit{i}) If all field modes are treated as classical, we recover the microscopic semi-classical result and the polarization is strictly linear in N.  Additional effects are obtained if we also add coupling to quantum modes and account for them order by order.  (\textit{ii}) Cascading and local-field effects are generated by including the coupling to quantum modes to second order (representing exchange of photons between two molecules).  The two effects have the same microscopic origin, which is not clear from the semi-clssical derivations. (\textit{iii}) Forster resonant energy transfer will be obtained at fourth order in the quantum mode as will three-molecule processes.  Such higher-order effects depend on generalized response functions and are not obtainable from a semi-classical perspective.  We limit our discussion to (\textit{i}) and (\textit{ii}).  Including higher-order effects is straightforward.
\par
We focus on a heterodyne detection signal in which the impinging field modes are all in a coherent (classical) state and take the field-matter coupling to be dipolar so that the interaction Hamiltonian is given by 
\[\hat{H}_{int}=-\int d\mathbf{r}\hat{\mathbf{E}}(\mathbf{r},t)\cdot\hat{\mathbf{V}}(\mathbf{r},t)\]
and the dipole operator is the sum of the dipole operators for each molecule $\hat{\mathbf{V}}(\mathbf{r},t)=\sum_a\hat{\mathbf{V}}_a(t)\delta(\mathbf{r}-\mathbf{r}_a)$. The electric field operator is partitioned into the sum of the classical  and vacuum modes:
\begin{align}\label{eq:Eops}
&\hat{\mathbf{E}}(\mathbf{r},t)=\sum_i\mathbf{E}_i(\mathbf{r},t)+\hat{\mathbf{E}}_v(\mathbf{r},t) \\ \notag
&\mathbf{E}_i(\mathbf{r},t)=\sum_{\zeta_i=\pm 1}\mathbf{\epsilon}_i\int\frac{d\omega_i}{2\pi}\mathcal{E}^{\zeta_i}_i(\omega_i)e^{i\zeta_i(\mathbf{k}_i\cdot\mathbf{r}-\omega_it)} \\ \notag
&\hat{\mathbf{E}}_v(\mathbf{r},t)=\sum_{\mathbf{k}_v\lambda}\sqrt{\frac{2\pi\hbar\omega_V}{\mathcal{V}}}\mathbf{\epsilon}^{(\lambda)}(\hat{\mathbf{k}}_v)\lbrace e^{i(\mathbf{k}_v\cdot\mathbf{r}-\omega_vt)}\hat{a}_{\mathbf{k}_v,\lambda}+e^{-i(\mathbf{k}_v\cdot\mathbf{r}-\omega_vt)}\hat{a}^{\dagger}_{\mathbf{k}_v,\lambda}\rbrace
\end{align}
Here, $\mathcal{E}^{\zeta_i}_i(t)$ is the temporal envelope of the $i$th pulse (or the conjugate for $\zeta_i=-1$) and $\mathbf{\epsilon}_i$ is its polarization vector. $\mathcal{V}$ is the quantization volume, $\lambda$ indexes the polarization of the vacuum mode and $\hat{a}^{(\dagger)}_{\mathbf{k}_v,\lambda}$ are vacuum mode annihilation (creation) operators. We begin with the Superoperator expression for the heterodyne signal derived from the rate of change of the photon number operator in the detected mode ($\langle \frac{d}{dt}N_s\rangle$) \cite{roslyak2010unified}:
\begin{equation}\label{eq:Sdef}
S=\frac{2}{\hbar }\Im\left\{\int dtd\mathbf{r}Tr\left[{\mathcal T}\hat{\mathbf{V}}_L\left({\mathbf r},t\right)\cdot\hat{\mathbf{E}}^{\dagger }_{sL}\left({\mathbf r},t\right)e^{\frac{-i}{\hbar }\int^t_{-\infty }{\hat{H}_{int-}(\tau )d\tau }}\rho \left(-\infty\right)\right]\right\}
\end{equation}
The symbol $\Im$ stands for the imaginary part and the subscript $L$ ($R$) on an operator indicates its action from the left (right).  For brevity, we also define the linear combinations:
\begin{equation}
\hat{O}_{-}=\hat{O}_L-\hat{O}_R
\end{equation}
\begin{equation}
\hat{O}_{+}=\frac{1}{2}(\hat{O}_L+ \hat{O}_R)
\end{equation}
If all field modes are classical, we recover the standard formulae for the heterodyne detected (stimulated emission) nonlinear signal in terms of the susceptibilities.  As described in references \cite{roslyak2010unified, dorfman2013nonlinear}, homodyne detected (spontaneously emitted) nonlinear signals arise from a $2$nd order interaction with a vacuum mode which is then detected.  Such processes are additive (i.e., the total signal for an aggregate of many molecules is simply the sum of the signals of the individual molecules).  These processes therefore scale linearly with $N$, the number of molecules in the sample.
\par
Like homodyne detected spontaneously emitted signals, cascades arise from interaction with a vacuum mode.  The difference is that in cascading processes, the final signal is still heterodyne detected and the quantum mode merely plays an intermediate role, causing an effective interaction between molecules that generates collective signals.  These cascading signals therefore scale as $N^2$ in the molecule number.  
\section{Vacuum-Mediated Interactions}
The first step in evaluating the signal from Eqn.~\ref{eq:Sdef} is to factorize the density matrix and separate the vacuum mode from the matter degrees of freedom:
\begin{equation}
\rho\left(-\infty\right)={\rho }_V(-\infty)\otimes {\rho}'\left(-\infty\right)
\end{equation}
where $\rho_V=\vert 0\rangle\langle 0\vert$ is the initial vacuum mode density matrix and $\rho'$ is the density matrix of the material.  To return $\rho_V$ to a population requires two interactions (one to excite a coherence between $\vert 0\rangle$ and $\vert 1\rangle$ and another to de-excite it) and so the lowest non-vanishing contribution is second order in the vacuum interactions. Cascading is a two-molecule process in which one of these interactions takes place on the molecule from which the signal is heterodyne detected (molecule $a$) and the other interaction takes place on a second molecule ($b$).  For a product of commuting operators (note that $\hat{\mathbf{E}}$ and $\hat{\mathbf{V}}$ act in separate spaces and therefore commute), $(AB)_-=A_-B_++A_+B_-$.  Since the $\mathbf{E}_{i}$ are c-numbers, $(\mathbf{E}_{i})_-=0$ and all interactions with classical fields are associated with a $\hat{\mathbf{V}}_-$.  Since $Tr[\hat{O}_-\rho]=0$ for any $\rho$, the final interaction on molecule $b$ must be the vacuum interaction ($(\hat{\mathbf{E}}_v)_-$ does not vanish).  Moreover, the vacuum interaction on molecule $b$ must come prior to that on molecule $a$ since otherwise the trace over the vacuum mode would vanish for this same reason. An important consequence of this reasoning is that the relevant correlation function for each molecule will be of the form $\langle V_+V_-\dots V_-\rangle$ (i.e. one "$+$" and several "$-$" indices) as in standard response functions. In this sense, nothing unusual happens to second-order in the vacuum modes.  It is worth noting however, that the classical correlation function of the vacuum mode $\langle E_{V+}E_{V+}\rangle$ turns out not to contribute while the cascading and local-field corrections are determined by $\langle E_{V+}E_{V-}\rangle$ which would vanish classically but is finite for the quantum vacuum.  This points to the fundamentally quantum nature of these signals and suggests that a broader variety of correlation functions may be accessible at higher order in the coupling to the vacuum modes.
\par
Below, we explore the corrections to the response, at various orders in the incoming fields, due to interactions between molecules mediated by second-order interaction with the vacuum mode.  We term these 2VMI corrections for shorthand and identify two relevent subsets.  As we will demonstrate, 2VMI corrections come as a sum of products of pairs of molecular hyperpolarizabilities (one each for molecules $a$ and $b$).  When one of these hyperpolarizabilities is first order (i.e. it is merely the linear polarizability), the process can be viewed as replacing the field that interacted with molecule $b$ by an effective field that then interacts with molecule $a$.  We term the corrections due to such processes ``local-field".  In the semi-classical treatments of local-field corrections, all incoming fields are replaced by effective fields while we only carry out this process with one incoming field.  The additional corrections are obtained at higher-order in the vacuum mode.  When neither of the hyperpolarizabilities is first-order, we term the process ``cascading".  
\par
\begin{figure}
\includegraphics{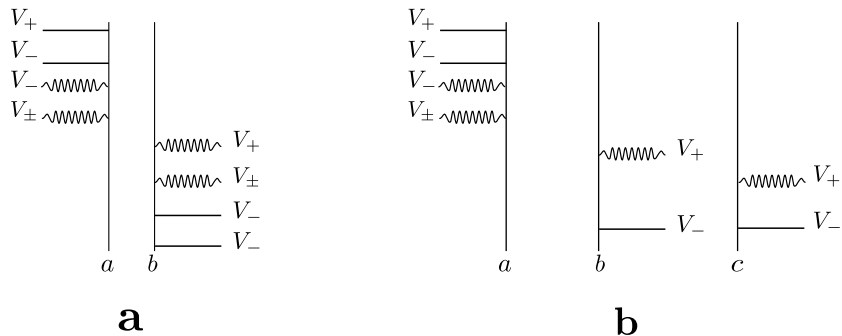}
\caption{$\mathbf{a}$: A two-molecule process involving either one or two vacuum modes. This can represent e.g., resonant energy transfer. $\mathbf{b}$: A three-molecule process in which molecules $b$ and $c$ each interact with the same vacuum mode which then interacts twice with molecule $a$. Both of these processes involve generalized response functions. Time progresses as one moves up the diagram and the vertical lines represent the density matrices of different molecules ($a$, $b$ and $c$).  Solid horizontal lines represent interactions with photons in the externally applied (laser) modes while the wavy lines represent interactions with vacuum-mode photons. Note that, since we work in the $\pm$ representation rather than the $L/R$, the diagrams do not distinguish between action on the ket or the bra.}
\label{fig:4vmi}
\end{figure}

As described above, the correlation functions that appear in the 2VMI corrections are all classical response functions of the $+-\dots-$ form. Beyond the second-order in the interaction with the vacuum, generalized material response functions of forms other than the usual $\langle V_+V_-\dots V_-\rangle$ begin to appear \cite{harbola2008superoperator}.  As an example, two processes that are fourth order in the vacuum modes (4VMI) are illustrated in Fig.~\ref{fig:4vmi} (though we do not treat them in this manuscript).  The most obvious process at this order (shown schematically in Fig.~(\ref{fig:4vmi}.a)) involves two vacuum interactions each at two different molecules and can represent resonant energy transfer.  A second process (Fig.~(\ref{fig:4vmi}.b)) involves three molecules and a single vacuum mode (the equivalent process with two vacuum modes will involve only ordinary response functions).  Since only the interactions with vacuum modes may be associated with a $\hat{\mathbf{V}}_+$, such higher-order processes access a broader array of material response functions.  Specifically, the processes depicted in Fig.~(\ref{fig:4vmi}) yield terms proportional to the non-classical correlation functions $\gamma_{+-+-}$ and $\gamma_{+--+}$.  Such processes are missed by the semi-classical approximation and are of potential interest particularly for harmonic systems in which the ordinary response function $\gamma_{+---}$ vanishes and variations from this are easier to detect.

\subsection{Local Field Corrections to the Linear Response}
\begin{figure}
\includegraphics{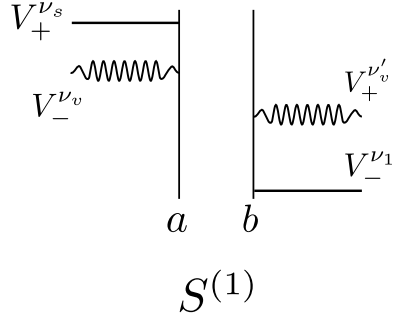}
\caption{Local field correction to the first-order signal. Time progresses as one moves up the diagram and the left and right vertical lines represent the density matrices of molecules $a$ and $b$ respectively.}
\label{fig:firstorder}
\end{figure}
With the above general considerations in mind, we begin by evaluating the 2VMI corrections to the heterodyne-detected, first-order signal.  At this order, there are two classical field interactions and two vacuum ineractions.  The only possible process to this order is shown schematically in Fig.~\ref{fig:firstorder} which gives
\begin{align}\label{eq:firstorder}
S^{(1)}(\Omega_s,\Omega_1)=&\frac{-4}{(2\pi)^2\hbar^4}\Im\bigg[\sum_{a,b}\sum_{\nu_i\zeta_i}\int d\omega_sd\omega_1\mathcal{E}^{\nu_s\dagger}_s(\omega_s)\mathcal{E}^{\nu_1,\zeta_1}_1(\omega_1)e^{i(\zeta_1\mathbf{k}_1\cdot\mathbf{r}_b-\mathbf{k}_s\cdot\mathbf{r}_a)}\\ \notag
&\mathcal{D}^{\nu_v\nu'_v}_{ab}(\zeta_1\omega_1){}^{(a)}\alpha_{+-}^{\nu_s\nu_v}(\zeta_1\omega_1) ^{(b)}\alpha_{+-}^{\nu'_v\nu_1}(\zeta_1\omega_1)\delta(\zeta_1\omega_1-\omega_s)\bigg].
\end{align}
The molecules are indexed by $a,b$ while the $\nu_i$ run over cartesian coordinates and the $\zeta_i$ tracks the hermiticity of each interaction. Note that we have written the signal as a function of $\Omega_s$ and $\Omega_1$, the central frequencies of the detection pulse and the interacting pulse.  More generally, the signal depends on all the parameters that define the pulse envelopes. The tensor $\mathcal{D}^{\nu_v\nu'_v}_{ab}(\omega)$ is defined in the appendix and accounts for the effects of the sample geometry. For a two-molecule sample, the near-field contribution goes as $\mathcal{D}^{\nu_v\nu'_v}_{ab}(\omega)\sim (\delta_{\nu_v\nu'_v}-3\hat{r}_{\nu_v}\hat{r}_{\nu'_v})r_{ab}^{-3}$.  In the case of an infinite homogeneous medium, we obtain instead the phase-matching condition via $\delta(\mathbf{k}_s-\mathbf{k}_v)\delta(\mathbf{k}_v-\mathbf{k}_1)$.  This gives the 2VMI corrections to the first-order heterodyne signal.  This is the same as the local-field corrections to first order in external fields since no cascading processes are possible at this order. 
\par
The underlying physical process for these corrections to the linear signal is the same as for the radiation-induced intermolecular energy shift of two molecules (or optical binding energy) with the signal mode the same as the applied external (i.e., $s=1$) \cite{thirunamachandran1980intermolecular, salam}.  The difference is that, since we examine $\langle \frac{d}{dt}N_s\rangle$ instead of $\Delta E$, the imaginary rather than the real part of the geometric coupling tensor ($\mathcal{D}^{\nu_v\nu'_v}_{ab}(\omega)$) is relevant. Note that, since we consider molecules with no permanent dipole, we only recover the dynamic and not the static contribution. 

\subsection{Local Field Corrections to the Second-Order Response}
\begin{figure}
\includegraphics[width=\textwidth]{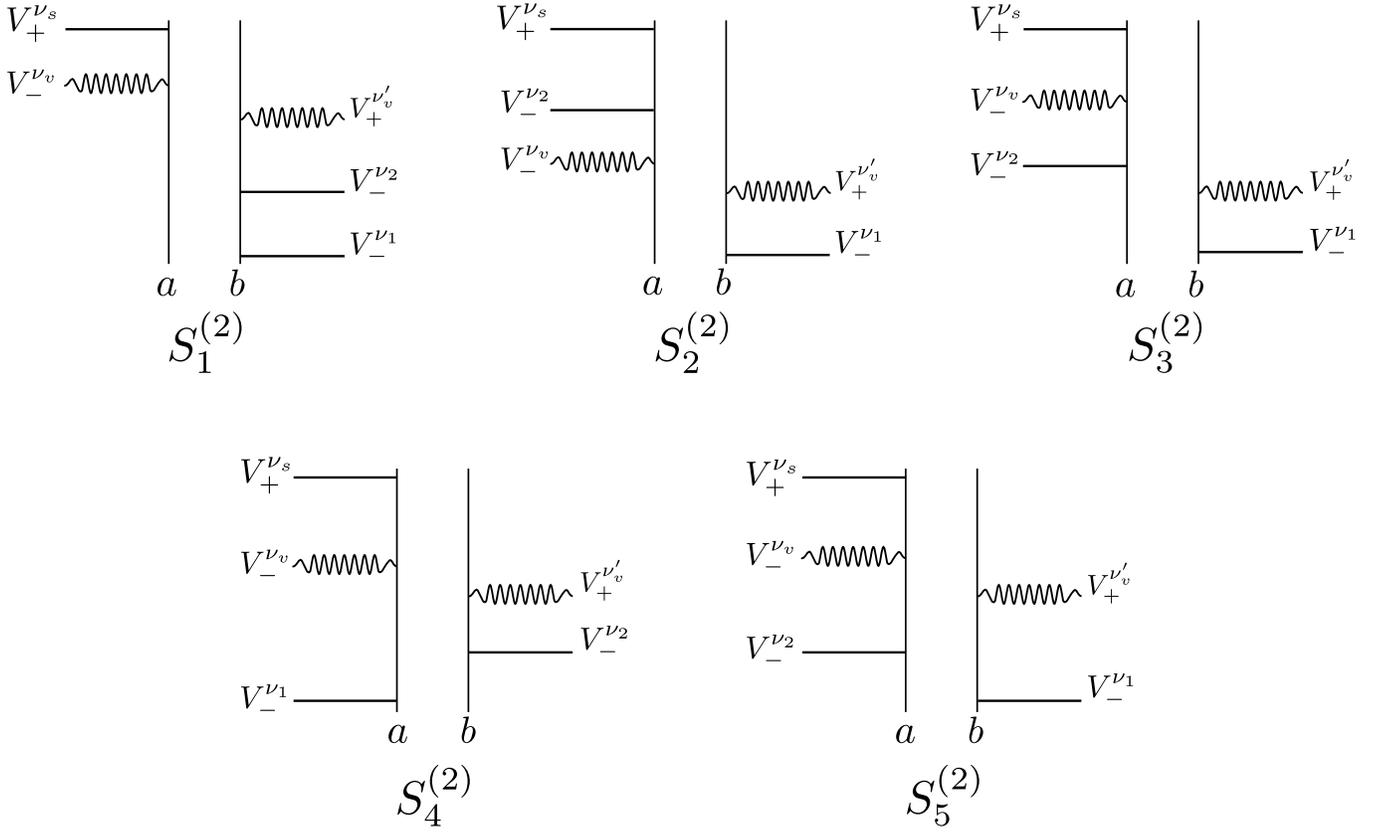}
\caption{Local field corrections to the second-order signal.  Diagrams 2, 3 and 5 can be combined since they all have molecule $b$ interacting with field $E_1$.}
\label{fig:secorder}
\end{figure}
Moving to second order, we acquire another interaction with external, classical fields.  The vacuum mode always has freedom to interact anytime after the first interaction in the series of classical fields but the vacuum interaction on molecule $b$ must be the final interaction for that molecule (since all other dipole operators on $b$ are $V_-$).  Assuming time-ordered pulses (pulse 1 comes before pulse 2 etc.), there are five possible orders of interactions as shown schematically by the diagrams in Fig.~\ref{fig:secorder}.  To exhaust all diagrams, we then permute $\lbrace 1,2\rbrace$ to give the other 5 diagrams.  The analysis of the linear response was carried out in the frequency representation.  For higher orders, the diagrams are more easily combined in the time representation and we proceed in this manner in the appendix. The result is most compactly expressed by defining a total response function:
\begin{align}
S^{(2)}(T_s,T_2,T_1)=\Im\bigg[\frac{4\pi}{\hbar^5}\sum_{ab}\sum_{\nu_i\zeta_i}\int d\tau_sd\tau_2d\tau_1d\tau_vE_s^{\nu_s\dagger}(\tau_s)E_2^{\nu_2\zeta_2}(\tau_2)E_1^{\nu_1\zeta_1}(\tau_1)\mathcal{R}_{ab}^{s21}(\tau_s,\tau_2,\tau_1,\tau_v)\bigg]
\end{align}
\begin{align}\label{eq:RTsecond}
\mathcal{R}_{ab}^{s21}(\tau_s,\tau_2,\tau_1,\tau_v)=&\sum_{\nu_v\nu'_v}\mathcal{C}_{ab}^{\nu_v\nu'_v}\bigg[e^{i(\zeta_2\mathbf{k}_2-\mathbf{k}_s)\cdot \mathbf{r}_a}e^{i\zeta_1\mathbf{k}_1\cdot\mathbf{r}_b}{}^{(a)}\beta_{+--}^{\nu_s\nu_2\nu_v}(\tau_s,\tau_2,\tau_v){}^{(b)}\alpha_{+-}^{\nu'_v\nu_1}\left(\tau_v-\frac{r_{ab}}{c},\tau_1\right)\\ \notag
&+e^{i(\zeta_1\mathbf{k}_1-\mathbf{k}_s)\cdot \mathbf{r}_a} e^{i\zeta_2\mathbf{k}_2\cdot\mathbf{r}_b}{}^{(a)}\beta_{+--}^{\nu_s\nu_1\nu_v}(\tau_s,\tau_1,\tau_v){}^{(b)}\alpha_{+-}^{\nu'_v\nu_1}\left(\tau_v-\frac{r_{ab}}{c},\tau_2\right)\\ \notag
&+e^{-i\mathbf{k}_s\cdot \mathbf{r}_a}e^{i(\zeta_1\mathbf{k}_1+\zeta_1\mathbf{k}_1)\cdot\mathbf{r}_b}{}^{(a)}\alpha_{+-}^{\nu_s\nu_v}(\tau_s,\tau_v){}^{(b)}{\beta}_{+--}^{\nu'_v\nu_2\nu_1}\left(\tau_v-\frac{r_{ab}}{c},\tau_2,\tau_1\right)
\end{align}
Where the superscripts $s21$ on $\mathcal{R}_{ab}$ indicate the dependence on $\nu_s, \nu_2,\nu_1,\zeta_2,\zeta_1$ and $\mathcal{C}_{ab}^{\nu_v\nu'_v}$ is defined in the appendix. We see that the time-representation yields compact expressions for the signal and that this approach reproduces the semi-classical result of a product of lower-order (hyper)polarizabilities. Note that the $T_i$ stand for the central time of the temporal pulse envelopes but that the signal generally depends on all parameters defining the pulse envelopes.  When a series of temporally non-overlapping pulses impinges on a sample, the pulses that interact with molecule $b$ are delayed from reaching molecule $a$ (on average, by the sum of the coherence decay time of molecule $b$ and the travel time between molecules).  If the coherence decay time of molecule $b$ is long relative to the distance between pulses, the order of interactions can switch.  That is, 2VMI corrections can scramble the time-ordering of applied pulses (this is observed at third-order in external fields in references \cite{DantusCascaded, cundiff2002time}).  To illustrate how the signal may be recast so as to highlight this scrambling, we examine the first term in Eqn.~\ref{eq:RTsecond} and define an effective field:
\begin{equation}
\tilde{E}^{\nu'_v}_1(\tau_v)=\sum_{\nu_1\zeta_1}\int d\tau_1E_1^{\nu_1\zeta_1}(\tau_1){}^{(b)}\alpha_{+-}^{\nu'_v\nu_1}\left(\tau_v-\frac{r_{ab}}{c},\tau_1\right)
\end{equation}
This permits the first term of Eqn.~\ref{eq:RTsecond} to be written as
\begin{align}
S^{(2)}_I=\Im\bigg[\frac{4\pi}{\hbar^5}\sum_{ab}\sum_{\nu_i\zeta_i}\int d\tau_sd\tau_2d\tau_1d\tau_vE_s^{\nu_s\dagger}(\tau_s)E_2^{\nu_2\zeta_2}(\tau_2)\tilde{E}^{\nu'_v}_1(\tau_v){}^{(a)}\beta_{+--}^{\nu_s\nu_2\nu_v}(\tau_s,\tau_2,\tau_v)\mathcal{C}_{ab}^{\nu_v\nu'_v}e^{i(\zeta_2\mathbf{k}_2-\mathbf{k}_s)\cdot \mathbf{r}_a}e^{i\zeta_1\mathbf{k}_1\cdot\mathbf{r}_b}\bigg]
\end{align}
From this form, it is clear that we may view molecule $b$ as generating an effective field the details of which are dependent on the shape of the impinging pulse as well the response of molecule $b$.  This effective field may therefore interact with molecule $a$ after the corresponding external field has already passed.  
\par
Having performed the analysis in the time domain (see appendix for details), we may change to the frequency domain and give the resulting expressions for completeness.  This is accomplished by substituting the electric field time-envelopes for their Fourier transforms and results in:
\begin{align}
S^{(2)}(\Omega_s,\Omega_2,\Omega_1)=\Im\bigg[\frac{1}{\pi\hbar^5}\sum_{ab}\sum_{\nu_i\zeta_i}\int d\omega_sd\omega_2d\omega_1\mathcal{E}_s^{\nu_s\dagger}(\omega_s)\mathcal{E}_2^{\nu_2\zeta_2}(\omega_2)\mathcal{E}_1^{\nu_1\zeta_1}(\omega_1)\delta(\zeta_1\omega_1+\zeta_2\omega_2-\omega_s)\mathcal{R}_{ab}^{s21}(\omega_s,\omega_2,\omega_1)\bigg]
\end{align}
\begin{align}
\mathcal{R}_{ab}^{s21}(\omega_s,\omega_2,\omega_1)=&\sum_{\nu_v\nu'_v}\bigg[e^{i(\zeta_2\mathbf{k}_2-\mathbf{k}_s)\cdot \mathbf{r}_a}e^{i\zeta_1\mathbf{k}_1\cdot\mathbf{r}_b}\mathcal{D}_{ab}^{\nu_v\nu'_v}(\zeta_1\omega_1){}^{(a)}\beta_{+--}^{\nu_s\nu_2\nu_v}(\zeta_1\omega_1,\zeta_2\omega_2){}^{(b)}\alpha_{+-}^{\nu'_v\nu_1}(\zeta_1\omega_1)\\ \notag
&+e^{i(\zeta_1\mathbf{k}_1-\mathbf{k}_s)\cdot \mathbf{r}_a} e^{i\zeta_2\mathbf{k}_2\cdot\mathbf{r}_b}\mathcal{D}_{ab}^{\nu_v\nu'_v}(\zeta_2\omega_2){}^{(a)}\beta_{+--}^{\nu_s\nu_1\nu_v}(\zeta_1\omega_1,\zeta_2\omega_2){}^{(b)}\alpha_{+-}^{\nu'_v\nu_2}(\zeta_2\omega_2)\\ \notag
&+e^{-i\mathbf{k}_s\cdot \mathbf{r}_a}e^{i(\zeta_1\mathbf{k}_1+\zeta_1\mathbf{k}_1)\cdot\mathbf{r}_b}\mathcal{D}_{ab}^{\nu_v\nu'_v}(\zeta_1\omega_1+\zeta_2\omega_2){}^{(a)}\alpha_{+-}^{\nu_s\nu_v}(\zeta_1\omega_1+\zeta_2\omega_2){}^{(b)}{\beta}_{+--}^{\nu'_v\nu_2\nu_1}(\zeta_1\omega_1,\zeta_2\omega_2)
\end{align}

\subsection{Quantum Field Corrections to the Third- and Fifth-Order Signals; Cascading}   
\begin{figure}
\includegraphics[width=\textwidth]{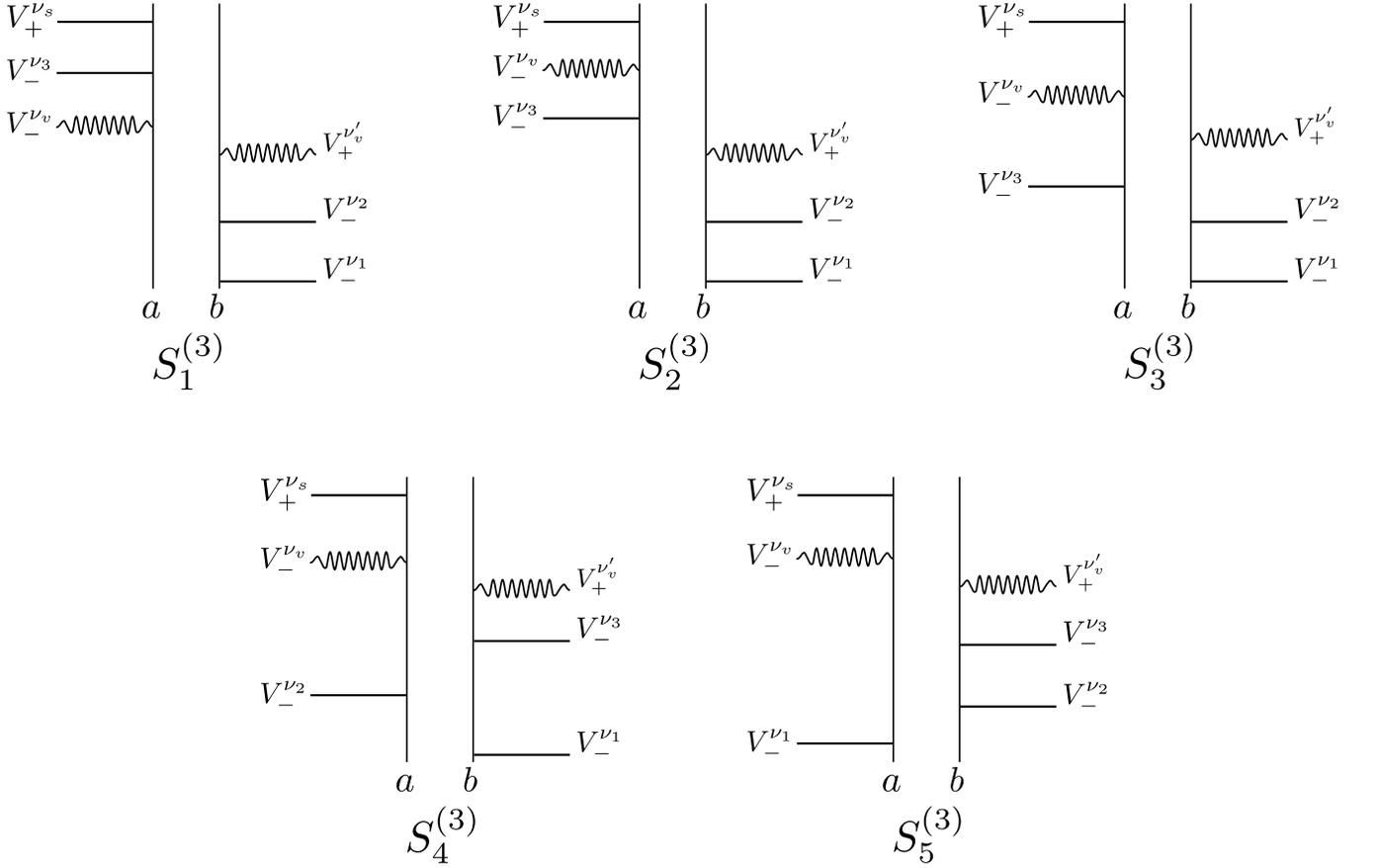}
\caption{Cascading contributions to the third-order signal.}
\label{fig:thirdorder}
\end{figure}
The total number of diagrams (before accounting for permutations of the interaction order) that contribute to order $n$ in external fields is
\begin{align}
N_D=\sum_m^n(2^m-1)(n-m+1)
\end{align}
where $N_D$  is the number of diagrams.  Third order thus contains 16 possible diagrams and most of these are of the local field type previously encountered thus far.  From this point forward we will focus on the cascading diagrams, defined as those in which the result is expressed as a product of lower-order correlation functions of the two molecules.  The total number of equal-order cascading diagrams (those cascading diagrams in which each hyperpolarizability is of the same order) that contribute at $n$th order in external fields is
\begin{align}\label{eq:cascnum}
N_{EOD}=\sum_m^n {m\choose \frac{n+1}{2}}(n-m+1)
\end{align}
and there are thus only 5 equal-order cascading diagrams at third order (shown in Fig.~\ref{fig:thirdorder}).  Note that these will appear only at odd orders and third is the first order for which they appear.  Permuting the order of field interactions generates 25 additional diagrams resulting in 30 total.  These diagrams carry various step functions and they may be combined along the same lines as shown in the previous section. The cascading response function for third order is thus:
\begin{align}
\mathcal{R}_{ab}^{s321}(\tau_s,\tau_3,\tau_2,\tau_1,\tau_v)=&\sum_{\nu_v\nu'_v}\mathcal{C}_{ab}^{\nu_v\nu'_v}\bigg[e^{i(\zeta_3\mathbf{k}_3-\mathbf{k}_s)\cdot \mathbf{r}_a}e^{i(\zeta_1\mathbf{k}_1+\zeta_2\mathbf{k}_2)\cdot\mathbf{r}_b}{}^{(a)}\beta_{+--}^{\nu_s\nu_3\nu_v}(\tau_s,\tau_3,\tau_v){}^{(b)}\beta_{+--}^{\nu'_v\nu_2\nu_1}\left(\tau_v-\frac{r_{ab}}{c},\tau_2,\tau_1\right)\\ \notag
&+e^{i(\zeta_3\mathbf{k}_3-\mathbf{k}_s)\cdot \mathbf{r}_a}e^{i(\zeta_1\mathbf{k}_1+\zeta_2\mathbf{k}_2)\cdot\mathbf{r}_b}{}^{(a)}\beta_{+--}^{\nu_s\nu_v\nu_2}(\tau_s,\tau_v,\tau_2){}^{(b)}\beta_{+--}^{\nu'_v\nu_3\nu_1}\left(\tau_v-\frac{r_{ab}}{c},\tau_3,\tau_1\right)\\ \notag
&+e^{i(\zeta_1\mathbf{k}_1-\mathbf{k}_s)\cdot \mathbf{r}_a}e^{i(\zeta_2\mathbf{k}_2+\zeta_3\mathbf{k}_3)\cdot\mathbf{r}_b}{}^{(a)}\beta_{+--}^{\nu_s\nu_v\nu_1}(\tau_s,\tau_v,\tau_1){}^{(b)}\beta_{+--}^{\nu'_v\nu_3\nu_2}\left(\tau_v-\frac{r_{ab}}{c},\tau_3,\tau_2\right)
\end{align}
or, in the frequency representation
\begin{align}\notag
\mathcal{R}_{ab}^{s321}(\omega_s,\omega_3,\omega_2,\omega_1,\tau_v)=&\sum_{\nu_v\nu'_v}\bigg[e^{i(\zeta_3\mathbf{k}_3-\mathbf{k}_s)\cdot \mathbf{r}_a}e^{i(\zeta_1\mathbf{k}_1+\zeta_2\mathbf{k}_2)\cdot\mathbf{r}_b}\mathcal{D}_{ab}^{\nu_v\nu'_v}(\Omega_{12}){}^{(a)}\beta_{+--}^{\nu_s\nu_3\nu_v}(\zeta_3\omega_3,\Omega_{12}){}^{(b)}\beta_{+--}^{\nu'_v\nu_2\nu_1}(\zeta_2\omega_2,\zeta_1\omega_1)\\ \notag
&+e^{i(\zeta_3\mathbf{k}_3-\mathbf{k}_s)\cdot \mathbf{r}_a}e^{i(\zeta_1\mathbf{k}_1+\zeta_2\mathbf{k}_2)\cdot\mathbf{r}_b}\mathcal{D}_{ab}^{\nu_v\nu'_v}(\Omega_{13}){}^{(a)}\beta_{+--}^{\nu_s\nu_v\nu_2}(\zeta_2\omega_2,\Omega_{13}){}^{(b)}\beta_{+--}^{\nu'_v\nu_3\nu_1}(\zeta_3\omega_3,\zeta_1\omega_1)\\ \notag
&+e^{i(\zeta_1\mathbf{k}_1-\mathbf{k}_s)\cdot \mathbf{r}_a}e^{i(\zeta_2\mathbf{k}_2+\zeta_3\mathbf{k}_3)\cdot\mathbf{r}_b}\mathcal{D}_{ab}^{\nu_v\nu'_v}(\Omega_{23}){}^{(a)}\beta_{+--}^{\nu_s\nu_v\nu_1}(\zeta_1\omega_1,\Omega_{23}){}^{(b)}\beta_{+--}^{\nu'_v\nu_3\nu_2}(\zeta_3\omega_3,\zeta_2\omega_2)\bigg]
\end{align}
where we have used the shorthand $\Omega_{ij}\equiv\zeta_i\omega_i+\zeta_j\omega_j$
\par
\begin{figure}
\includegraphics[width=\textwidth]{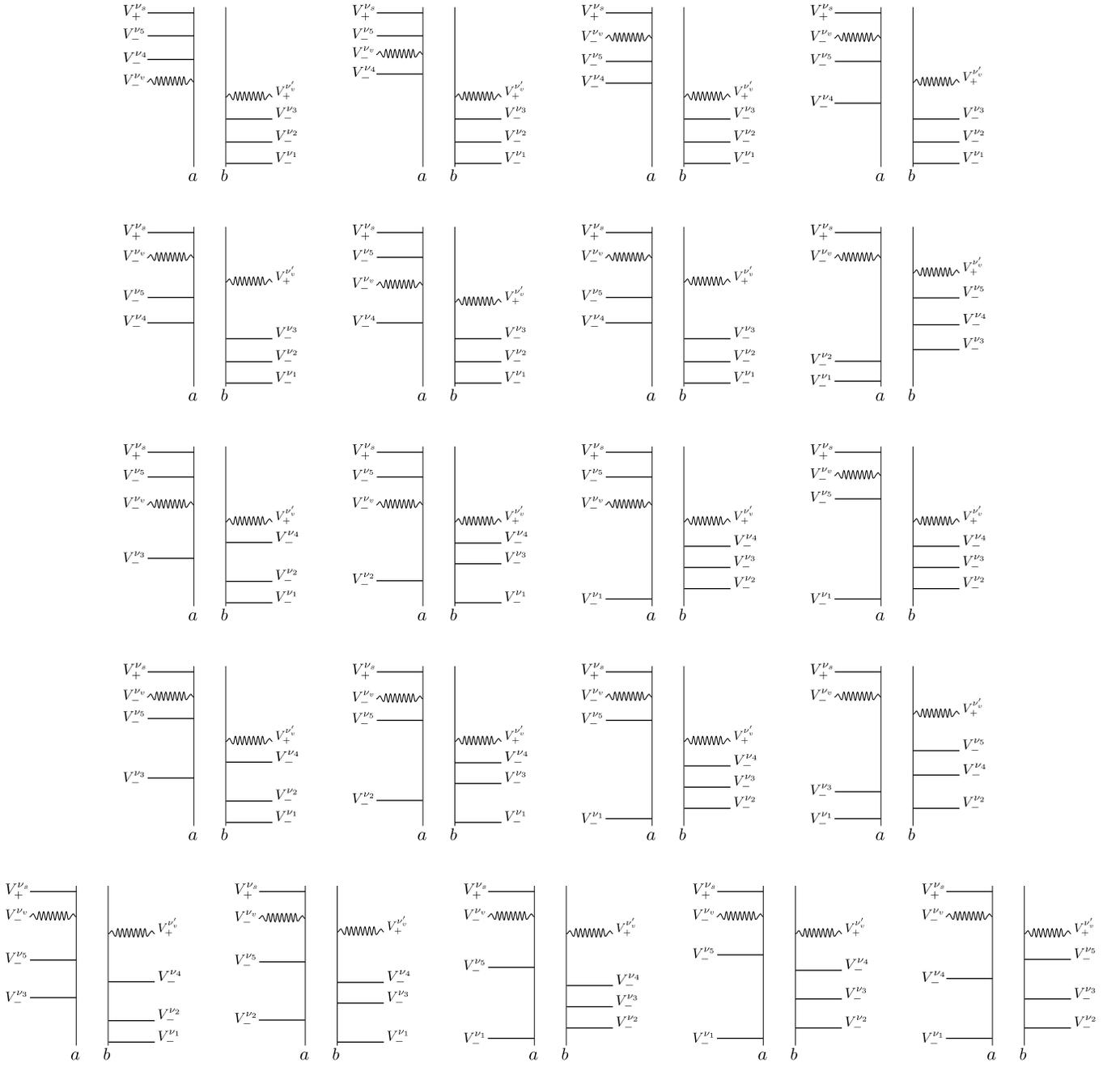}
\caption{Cascading Diagrams for Fifth Order}\label{fig:fifthorder}
\end{figure}

As per Eqn.~\ref{eq:cascnum}, there are 21 diagrams (shown in Fig.~\ref{fig:fifthorder}) at fifth order in the external fields.  Expressions for these diagrams follow in the same manner as before.  Considering only equal-order cascading contributions, the signal will come as a sum of products of ${}^{(a)}\gamma{}^{(b)}\gamma$. Because the even-order response generally vanishes (for material samples possessing inversion symmetry), this is the first cascading term that we might expect to significantly alter the signal.  Indeed, it was found that fifth-order Raman processes are generally dominated by cascading $\chi^{(3)}$ signals \cite{blank1999, zhao2011}.   

\section{Conclusions}
Expressions for cascading signals are traditionally derived semi-classically by creating a polarization and propagating with Maxwell's equations \cite{grimberg2002ultrafast} while local-field effects are included by considering a cavity within a homogenous dielectric medium.  Besides obscuring these effects' common origin in vacuum-mediated interactions between particles, one cannot know a priori whether or not all relevent effects have been included since the macroscopic semi-classical approach is fundamentally ad hoc.  In this paper, we present a microscopic quantum field derivation that unifies local-field and cascading effects and is systematic, leading to a general class of corrections due to vacuum interactions.  We find that the correction to the macroscopic response due to second-order interaction with the vacuum can be expressed as a sum of products of pairs of molecular response functions.  This treatment leads to local-field and cascading effects that can alter the signal appreciably (e.g. by overwhelming a direct process of greater interest \cite{blank1999, zhao2011}, scrambling the time-ordering of externally applied pulses \cite{DantusCascaded, cundiff2002time}, or altering the magnitude of the response \cite{louie1975local, Hache:86}). Although this result agrees with that of the macroscopic semi-classical approach and all correlation functions are of the $+-\dots-$ form (that of a standard response function from semi-classical theory), the approach immediately suggests processes of higher order in vacuum interactions. Two 4th order examples are Forster resonant energy transfer (Fig.~(\ref{fig:4vmi}.a) and the three-molecule process shown in Fig.~(\ref{fig:4vmi}.b) Both of these processes will yield terms proportional to $\gamma_{+-+-}$ and $\gamma_{+--+}$ which never arise in the semi-classical approach.

\acknowledgments
The support of the Chemical Sciences, Geosciences, and Biosciences division, Office of Basic Energy Sciences, Office of Science, U.S. Department of Energy is gratefully acknowledged.  We also gratefully acknowledge the support of the National Science Foundation (grants CHE-1058791 and CHE-0840513), and the National Institutes of Health (Grant GM-59230).

\appendix
\section{Quantum-Field Corrections to the First-Order Response}
Using  Fig.~\ref{fig:firstorder}, we expand Eqn.~\ref{eq:Sdef} to second order in interactions with the vacuum mode and first order in external modes.  This gives
\begin{align}
&S^{(1)}=\frac{2(-i)^3}{\hbar^4}\Im\bigg[\sum_{a,b}\sum_{\nu_i\zeta_i}\int d\tau'_s\int_{-\infty}^{\tau'_s}d\tau_v\int_{-\infty}^{\tau_v}d\tau_v' \int_{-\infty}^{\tau_v'}d\tau'_1 \int\frac{d\omega_sd\omega_1}{(2\pi)^2}\mathcal{E}^{\nu_s\dagger}_s(\omega_s)\mathcal{E}^{\nu_1,\zeta_1}_1(\omega_1)\\ \notag
&\times \langle V^{\nu_s}_+(\tau'_s)V^{\nu_v}_-(\tau_v)\rangle_a\langle V^{\nu'_v}_+(\tau'_v)V^{\nu_1}_-(\tau'_1)\rangle_b\sum_{\mathbf{k}_v,\lambda}\frac{2\pi\omega_v}{\mathcal{V}} e^{-i(\mathbf{k}_s\cdot\mathbf{r}_a-\omega_s\tau'_s)}e^{i\zeta_1(\mathbf{k}_1\cdot\mathbf{r}_-\omega_1\tau'_1)} \mathbf{\epsilon}_{\nu_v}^{(\lambda)}(\hat{\mathbf{k}}_v)\mathbf{\epsilon}_{\nu'_v}^{(\lambda)}(\hat{\mathbf{k}}_v)\\ \notag
&\times \lbrace \langle a_+a_-^\dagger\rangle_ve^{i(\mathbf{k}_v\cdot(\mathbf{r}_a-\mathbf{r}_b)-\omega_v(\tau_v-\tau'_v))}+\langle a^\dagger_+a_-\rangle_ve^{-i(\mathbf{k}_v\cdot(\mathbf{r}_a-\mathbf{r}_b)-\omega_v(\tau_v-\tau'_v))}\rbrace\bigg]
\end{align}
Here, we have used the definition of the dipole operator and the expressions in Eq.~(\ref{eq:Eops}) and factored the trace into a product of traces over molecule $a$, molecule $b$, and the vacuum mode.  The factors of $\left(\frac{-i}{\hbar}\right)^3$ come from the three interections in the expansion and the $\nu_j$ are cartesian coordinates.  Since we do not work in the rotating wave, we must account for both hermiticities of each operator.  This is done explicitly for the vacuum mode above and is the source of the two different terms in braces.  For the classical modes, this will be handled with the $\zeta_i$ factors while it will be included only implicitly for the material dipole operators.  Although we could proceed with the $\tau$ variables and will do so for higher-order corrections, we will demonstrate how the analysis goes in the frequency domain with the correction to first-order.  To this end, we change variables to the time between interactions
\begin{equation}
\tau'_s-\tau_v\equiv t_3,~~\tau_v-\tau'_v\equiv t_2,~~\tau'_v-\tau_1\equiv t_1
\end{equation}
and perform several simplifications (given in \cite{salam})
\begin{align}
&\sum_\lambda \mathbf{\epsilon}_{\nu_v}^{(\lambda)}(\hat{\mathbf{k}}_v)\mathbf{\epsilon}_{\nu'_v}^{(\lambda)}(\hat{\mathbf{k}}_v)=\delta_{\nu_{v}\nu'_v}-\hat{\mathbf{k}}_{v_{\nu_v}}\hat{\mathbf{k}}_{v_{\nu_v'}}\\ \notag
&\frac{1}{\mathcal{V}}\sum_{\mathbf{k}_v}\to\int\frac{d\omega_vd\Omega_v\omega_v^2}{(2\pi c)^3}\\ \notag
&\int d\Omega_v\left(\delta_{\nu_v\nu'_v}-\hat{\mathbf{k}}_{v_{\nu_v}}\hat{\mathbf{k}}_{v_{\nu'_v}}\right)e^{\pm i\mathbf{k}_v\cdot\mathbf{r}}= \left(-\nabla^2\delta_{\nu_v\nu'_v}+\nabla_{\nu_v}\nabla_{\nu'_v}\right)\frac{\sin{k_vr}}{k_v^3r}
\end{align}
resulting in the following form for the signal
\begin{align}
&S^{(1)}=\frac{-4i}{\hbar^4}\Im\bigg[\sum_{a,b}\sum_{\nu_i\zeta_i}\int_0^\infty dt_2\int d\omega_sd\omega_1\delta(\zeta_1\omega_1-\omega_s)\mathcal{E}^{\nu_s\dagger}_s(\omega_s)\mathcal{E}^{\nu_1,\zeta_1}_1(\omega_1)e^{i(\zeta_1\mathbf{k}_1\cdot\mathbf{r}_b-\mathbf{k}_s\cdot\mathbf{r}_a)}\langle V_+^{\nu_s}G(\zeta_1\omega_1)V_-^{\nu_v}\rangle_a\\ \notag
&\times\langle V_+^{\nu'_v}G(\zeta_1\omega_1)V_-^{\nu_1}\rangle_b\int\frac{d\omega_v}{(2\pi)^3}\left(-\nabla^2\delta_{\nu_v\nu'_v}+\nabla_{\nu_v}\nabla_{\nu'_v}\right)\frac{\sin{k_vr}}{r}\lbrace e^{i(\zeta_1\omega_1-\omega_v)t_2}-e^{i(\zeta_1\omega_1+\omega_v)t_2}\rbrace\bigg]
\end{align}
To arrive at the above, we have also changed from time-dependent operators to a Green's function representation:
\begin{align}
&G(t)\equiv -i \theta(t)e^{-i\hat{H}_{0-}(t)}\\ \notag
&G(\omega)=\int dt G(t)e^{i\omega t}=\frac{1}{\omega-\hat{H}_{0-}+i\eta}
\end{align}
with $\eta$ a positive infinitessimal.  We may now carry out the $t_2$ and $\omega_v$ integrations via:
\begin{align}
&\int d\omega_v\sin{\left(\omega_v\frac{r_{ab}}{c}\right)}\int_0^{\infty}dt_2(-i)\bigg\lbrace  e^{i(\zeta_1\omega_1-\omega_v)t_2}-e^{i(\zeta_1\omega_1+\omega_v)t_2}\bigg\rbrace=\\ \notag
\int d&\omega_v \sin{\left(\omega_v\frac{r_{ab}}{c}\right)}\bigg\lbrace\frac{1}{\zeta_1\omega_1-\omega_v+i\eta}-\frac{1}{\zeta_1\omega_1+\omega_v+i\eta}\bigg\rbrace=-2\pi e^{i\zeta_1\omega_1\frac{r_{ab}}{c}}
\end{align}
In the above integration over $d\omega_v$, the first term has poles in the upper half plane (UHP) and the second has poles in the lower half plane (LHP).  Because $\sin{x}=\frac{e^{ix}-e^{-ix}}{2i}$, the first term picks up the positive exponential and the second picks up the negative exponential.  Since the sign of the pole also changes between these two terms, the sign changes cancel and the both terms contribute a positive exponential.  To simplify the resulting expression while maintaining generality, we introduce 
\begin{equation}
\mathcal{D}^{\nu_v\nu'_v}_{ab}(\zeta_1\omega_1)=\left(-\nabla^2\delta_{\nu_v\nu'_v}+\nabla_{\nu_v}\nabla_{\nu'_v}\right)\frac{e^{i\zeta_1\omega_1\frac{r_{ab}}{c}}}{r_{ab}}
\end{equation}
We will also define
\begin{equation}
^{(a,b)}\bar{\alpha}_{+-}^{\nu_i\nu_j}(\omega)\equiv\langle V_+^{\nu_i}G(\omega)V_-^{\nu_j}\rangle_{(a,b)}
\end{equation}
Since there is only one external field, this is the same as the usual first-order polarizability ${}^{(a,b)}\alpha_{+-}^{\nu_i\nu_j}(\omega)$ \cite{mukbook}.  The definition is therefore superfluous at first order but is prototypical of the time-ordered hyperpolarizabilities that appear in higher order diagrams.
Combining the above results gives Eqn.~\ref{eq:firstorder}.
\par
In the case that the sample consists merely of two isolated molecules, we use the identity
\begin{equation}
\left(-\nabla^2\delta_{\nu_v\nu'_v}+\nabla_{\nu_v}\nabla_{\nu'_v}\right)\frac{e^{i\omega r}}{r}=\frac{1}{r^3}\big[(\delta_{\nu_v\nu'_v}-3\hat{r}_{\nu_v}\hat{r}_{\nu'_v})(1-i\omega r)+(\delta_{\nu_v\nu'_v}-\hat{r}_{\nu_v}\hat{r}_{\nu'_v})\omega^2r^2\big]e^{i\omega r}
\end{equation}
and we see that there are terms proportional to $r^{-1}$, $r^{-2}$ and $r^{-3}$ the last of which is dominant in the near-field regime.  In the case of an infinite homogeneous and isotropic medium (as is used in the macroscopic derivation of cascading terms \cite{mukbook, boydnonlinear}), the analysis is easier if one performs the summations over molecules before the integral over vacuum modes.  Since the molecules are identical, they have the same molecular response functions and these can be taken out of the integration over the molecules:
\begin{equation}
\int d\mathbf{r}_ad\mathbf{r}_be^{i(-\mathbf{k}_s+\zeta_v\mathbf{k}_v)\cdot\mathbf{r}_a}e^{i(-\zeta_v\mathbf{k}_v+\zeta_1\mathbf{k}_1)\cdot\mathbf{r}_a}= 4\pi^2\delta(\mathbf{k}_s-\mathbf{k}_v)\delta(\mathbf{k}_v-\mathbf{k}_1)
\end{equation}
These delta functions enforce the same phase matching that the direct (in this case first-order) process possesses.

\section{Quantum-Field Corrections to the Second-Order Response}
The analysis of the linear response was carried out in the frequency representation.  For higher order signals, the diagrams are more easily combined in the time representation and we will proceed in this manner.  
\par
We consider first the three diagrams in which field $1$ interacts with molecule $b$, namely $S^{(2)}_2$, $S^{(2)}_3$, and $S^{(2)}_5$.  From the diagram we have 
\begin{align}
S^{(2)}_2=&\Im\bigg[\frac{2(-i)^4}{\hbar^5}\sum_{a,b}\sum_{\nu_i\zeta_i}\int d\tau_sd\tau_2d\tau_1d\tau_vd\tau'_v\theta(\tau_s-\tau_2)\theta(\tau_2-\tau_v)\theta(\tau_v-\tau_v')\theta(\tau'_v-\tau_1) E_s^{\nu_s\dagger}(\mathbf{r}_a,\tau_s)E_2^{\nu_2\zeta_2}(\mathbf{r}_a,\tau_2)E_1^{\nu_1\zeta_1}(\mathbf{r}_b,\tau_1)\\ \notag
&\times\left(-\nabla^2\delta_{\nu_v\nu'_v}+\nabla_{\nu_v}\nabla_{\nu'_v}\right)\langle V_+^{\nu_s}(\tau_s)V_-^{\nu_2}(\tau_2)V_-^{\nu_v}(\tau_v)\rangle_a\langle V_+^{\nu'_v}(\tau'_v)V_-^{\nu_1}(\tau_1)\rangle_b\frac{2\pi}{ir_{ab}}\lbrace\delta(\tau'_v-\tau_v-\frac{r_{ab}}{c})-\delta(\tau'_v-\tau_v+\frac{r_{ab}}{c})\rbrace\bigg]
\end{align}
where we have used
\begin{align}
\int d\omega_v\sin{\left(\omega_v\frac{r_{ab}}{c}\right)}\big[e^{-i\omega_v(\tau_v-\tau'_v)}-e^{i\omega_v(\tau_v-\tau'_v)}\big]=\frac{2\pi}{i}\big[\delta(\tau'_v-\tau_v-\frac{r_{ab}}{c})-\delta(\tau'_v-\tau_v+\frac{r_{ab}}{c})\big]
\end{align}
Since $\tau_v>\tau'_v$ (based on the general considerations given under section II above and explicitly enforced by the factor $\theta(\tau_v-\tau'_v)$) and $\frac{r_{ab}}{c}$ is inherently positive, only the second $\delta$-function contributes.
\begin{align}
S^{(2)}_2=\Im\bigg[&\frac{4\pi(-i)^3}{\hbar^5}\sum_{a,b}\sum_{\nu_i\zeta_i}\int d\tau_sd\tau_2d\tau_1d\tau_v\theta(\tau_s-\tau_2)\theta(\tau_2-\tau_v)\theta(\tau_v-\frac{r_{ab}}{c} -\tau_1) E_s^{\nu_s\dagger}(\mathbf{r}_a,\tau_s)E_2^{\nu_2\zeta_2}(\mathbf{r}_a,\tau_2)E_1^{\nu_1\zeta_1}(\mathbf{r}_b,\tau_1)\\ \notag
&\times\left(-\nabla^2\delta_{\nu_v\nu'_v}+\nabla_{\nu_v}\nabla_{\nu'_v}\right)\frac{1}{r_{ab}}\langle V_+^{\nu_s}(\tau_s)V_-^{\nu_2}(\tau_2)V_-^{\nu_v}(\tau_v)\rangle_a\langle V_+^{\nu'_v}(\tau_v-\frac{r_{ab}}{c})V_-^{\nu_1}(\tau_1)\rangle_b\bigg]
\end{align}
where we have dropped the factor $\theta(\frac{r_{ab}}{c})$ since this is always satisfied.  Substituting in the Green's functions and defining an effective field
\begin{align}
\tilde{E}^{\nu'_v}_1(\mathbf{r}_b,\tau_v-\frac{r_{ab}}{c})=\sum_{\nu_1\zeta_1}\int d\tau_1E_1^{\nu_1\zeta_1}(\mathbf{r}_b,\tau_1)\langle V_+^{\nu'_v}G(\tau_v-\frac{r_{ab}}{c}-\tau_1)V_-^{\nu_1}\rangle_b
\end{align}
and the tensor
\begin{equation}
\mathcal{C}^{\nu_v\nu'_v}_{ab}=\left(-\nabla^2\delta_{\nu_v\nu'_v}+\nabla_{\nu_v}\nabla_{\nu'_v}\right)\frac{1}{r_{ab}}
\end{equation}
for shorthand gives the following signal
\begin{align}
&S^{(2)}_2=\Im\bigg[\frac{4\pi}{\hbar^5}\sum_{a,b}\sum_{\nu_i\zeta_2}\int d\tau_sd\tau_2d\tau_vE_s^{\nu_s\dagger}(\mathbf{r}_a,\tau_s)E_2^{\nu_2\zeta_2}(\mathbf{r}_a,\tau_2)\mathcal{C}^{\nu_v\nu'_v}_{ab}{}^{(a)}\bar{\beta}^{\nu_s\nu_2\nu_v}_{+--}(\tau_s-\tau_2,\tau_2-\tau_v)\tilde{E}_1^{\nu'_v}(\mathbf{r}_b,\tau_v-\frac{r_{ab}}{c})\bigg]
\end{align}
where we have defined the material correlation function:
\begin{align}
{}^{(a)}\bar{\beta}^{\nu_i\nu_j\nu_k}_{+--}(t,t')\equiv\langle V_+^{\nu_i}G(t)V_-^{\nu_j}G(t')V_-^{\nu_k}\rangle_a
\end{align}
Following the same procedure with $S^{(2)}_3$ yields
\begin{align}
S^{(2)}_3=\Im\bigg[\frac{4\pi}{\hbar^5}\sum_{a,b}\sum_{\nu_i\zeta_2}&\int d\tau_sd\tau_2d\tau_vE_s^{\nu_s\dagger}(\mathbf{r}_a,\tau_s)E_2^{\nu_2\zeta_2}(\mathbf{r}_a,\tau_2)\theta\left(\tau_2-(\tau_v-\frac{r_{ab}}{c})\right)\\ \notag
&\times\mathcal{C}^{\nu'_v\nu_v}_{ab}\tilde{E}_1^{\nu'_v}(\mathbf{r}_b,\tau_v-\frac{r_{ab}}{c}){}^{(a)}\bar{\beta}^{\nu_s\nu_2\nu_v}_{+--}(\tau_s-\tau_v,\tau_v-\tau_2)\bigg]
\end{align}
This expression clearly has the same form as the above for $S^{(2)}_2$ except for the additional factor of $\theta\left(\tau_2-\tau_v+\frac{r_{ab}}{c}\right)$ which enforces the fact that the second pulse only has a time $\frac{r_{ab}}{c}$ to interact with molecule $a$ since it must do so before $\tau_v$ in this diagram.
\par
Lastly, we consider $S^{(2)}_5$.  Analysis of this diagram is only somewhat more subtle and after following the previous steps in analogy we arrive at
\begin{align}
S^{(2)}_5=\Im\bigg[\frac{4\pi(-i)^3}{\hbar^5}\sum_{a,b}\sum_{\nu_i\zeta_2}&\int d\tau_sd\tau_2d\tau_vd\tau_1E_s^{\nu_s\dagger}(\mathbf{r}_a,\tau_s)E_2^{\nu_2\zeta_2}(\mathbf{r}_a,\tau_2)E_1^{\nu_1\zeta_1}(\mathbf{r}_b,\tau_1)\theta(\tau_s-\tau_v)\theta\left(\tau_v-\frac{r_{ab}}{c}-\tau_2\right)\\ \notag
&\times\theta(\tau_2-\tau_1)\mathcal{C}^{\nu_v\nu'_v}_{ab}\langle V_+^{\nu_s}(\tau_s)V_-^{\nu_v}(\tau_v)V_-^{\nu_2}(\tau_2)\rangle_a\langle V_+^{\nu'_v}\left(\tau_v-\frac{r_{ab}}{c}\right)V_-^{\nu_1}(\tau_1)\rangle_b\bigg]
\end{align}
The presence of the factors $\theta\left(\tau_v-\frac{r_{ab}}{c}-\tau_2\right)$ and $\theta(\tau_2-\tau_1)$ means that a factor of $\theta\left(\tau_v-\frac{r_{ab}}{c}-\tau_1\right)$ is redundant and may be freely added.  We do so in order to express the correlation function of molecule $b$ in terms of the Green's function.  Additionally, the positivity of $\frac{r_{ab}}{c}$ allows us to add the redundant factor $\theta(\tau_v-\tau_2)$ so as to do the same for the correlation function of molecule $a$.  These considerations result in:
\begin{align}
S^{(2)}_5=\Im\bigg[\frac{4\pi}{\hbar^5}\sum_{a,b}\sum_{\nu_i\zeta_2}&\int d\tau_sd\tau_2d\tau_vE_s^{\nu_s\dagger}(\mathbf{r}_a,\tau_s)E_2^{\nu_2\zeta_2}(\mathbf{r}_a,\tau_2)\theta\left(\tau_v-\frac{r_{ab}}{c}-\tau_2\right)\\ \notag
&\times\theta(\tau_2-\tau_1)\mathcal{C}^{\nu_v\nu'_v}_{ab}\tilde{E}_1^{\nu'_v}(\mathbf{r}_b,\tau_v-\frac{r_{ab}}{c}){}^{(a)}\bar{\beta}^{\nu_s\nu_v\nu_2}_{+--}(\tau_s-\tau_v,\tau_v-\tau_2)\bigg]
\end{align}
A careful look at the diagram for $S_4^{(2)}$ reveals that the corresponding diagram with $1\leftrightarrow 2$ can, by these same steps, be brought to the form:
\begin{align}
S^{(2)}_4(1\leftrightarrow2)=\Im\bigg[\frac{4\pi}{\hbar^5}\sum_{a,b}\sum_{\nu_i\zeta_2}&\int d\tau_sd\tau_2d\tau_vE_s^{\nu_s\dagger}(\mathbf{r}_a,\tau_s)E_2^{\nu_2\zeta_2}(\mathbf{r}_a,\tau_2)\theta\left(\tau_v-\frac{r_{ab}}{c}-\tau_2\right)\\ \notag
&\times\theta(\tau_1-\tau_2)\mathcal{C}^{\nu_v\nu'_v}_{ab}\tilde{E}_1^{\nu'_v}(\mathbf{r}_b,\tau_v-\frac{r_{ab}}{c}){}^{(a)}\bar{\beta}^{\nu_s\nu_v\nu_2}_{+--}(\tau_s-\tau_v,\tau_v-\tau_2)\bigg]
\end{align}
which, is identical to the expression for $S_5^{(2)}$ except the factor of $\theta(\tau_2-\tau_1)$ has been replaced by $\theta(\tau_1-\tau_2)$. Since the positivity of the Green's function arguments has all been assured by their separate $\theta$-functions (which are part of the Green's function by definition) and no operators acting in the same space differ in chronological order between the two terms, we may combine these two results using the identity $\theta(x-x_0)+\theta(x_0-x)=1$ to obtain:
\begin{align}
S^{(2)}_4(1\leftrightarrow2)+S_5^{(2)}=\Im\bigg[\frac{4\pi}{\hbar^5}\sum_{a,b}\sum_{\nu_i\zeta_2}&\int d\tau_sd\tau_2d\tau_vE_s^{\nu_s\dagger}(\mathbf{r}_a,\tau_s)E_2^{\nu_2\zeta_2}(\mathbf{r}_a,\tau_2)\theta\left(\tau_v-\frac{r_{ab}}{c}-\tau_2\right)\\ \notag
&\times\mathcal{C}^{\nu_v\nu'_v}_{ab}\tilde{E}_1^{\nu'_v}(\mathbf{r}_b,\tau_v-\frac{r_{ab}}{c}){}^{(a)}\bar{\beta}^{\nu_s\nu_v\nu_2}_{+--}(\tau_s-\tau_v,\tau_v-\tau_2)\bigg]
\end{align}
This result differs from $S_3^{(2)}$ by $\theta\left(\tau_v-\frac{r_{ab}}{c}-\tau_2\right)\to\theta\left(\tau_2-(\tau_v-\frac{r_{ab}}{c})\right)$ and so we combine them as before.  The result will differ from $S_2^{(2)}$ only in the order of the interactions on the material correlation function for molecule $a$ and so we may combine all of these results to yield
\begin{align}
&S_2^{(2)}+S_3^{(2)}+S^{(2)}_4(1\leftrightarrow2)+S_5^{(2)}=\\ \notag
\Im\bigg[\frac{4\pi}{\hbar^5}\sum_{a,b}\sum_{\nu_i\zeta_2}\int d\tau_sd\tau_2d\tau_v&E_s^{\nu_s\dagger}(\mathbf{r}_a,\tau_s)E_2^{\nu_2\zeta_2}(\mathbf{r}_a,\tau_2)\mathcal{C}^{\nu_v\nu'_v}_{ab}\tilde{E}_1^{\nu'_v}(\mathbf{r}_b,\tau_v-\frac{r_{ab}}{c}){}^{(a)}\beta^{\nu_s\nu_v\nu_2}_{+--}(\tau_s,\tau_v,\tau_2)\bigg]
\end{align}
where we have used the time-ordered correlation function
\begin{align}
&{}^{(a)}\beta_{+--}^{\nu_i\nu_j\nu_k}(t_i,t_j,t_k)=\langle \mathcal{T}V_+^{\nu_i}(t_i)V_-^{\nu_j}(t_j)V_-^{\nu_k}(t_k)\rangle_a
\end{align}
It is now clear that, in an analogous fashion, we obtain
\begin{align}
&S_2^{(2)}(1\leftrightarrow2)+S_3^{(2)}(1\leftrightarrow2)+S^{(2)}_4+S_5^{(2)}(1\leftrightarrow2)=\\ \notag
\Im\bigg[\frac{4\pi}{\hbar^5}\sum_{a,b}\sum_{\nu_i\zeta_2}\int d\tau_sd\tau_2d\tau_v&E_s^{\nu_s\dagger}(\mathbf{r}_a,\tau_s)E_1^{\nu_1\zeta_1}(\mathbf{r}_a,\tau_1)\mathcal{C}^{\nu_v\nu'_v}_{ab}\tilde{E}_2^{\nu'_v}(\mathbf{r}_b,\tau_v-\frac{r_{ab}}{c}){}^{(a)}\beta^{\nu_s\nu_v\nu_1}_{+--}(\tau_s,\tau_v,\tau_1)\bigg]
\end{align}
The only remaining diagrams to consider are  
\begin{align}
S^{(2)}_1+S^{(2)}_1(1\leftrightarrow2)=\Im\bigg[\frac{4\pi}{\hbar^5}\sum_{a,b}\sum_{\nu_i}\int d\tau_sd\tau_vE_s^{\nu_s\dagger}(\mathbf{r}_a,\tau_s)\mathcal{C}^{\nu_v\nu'_v}_{ab}\tilde{E}_{21}^{\nu'_v}(\mathbf{r}_b,\tau_v-\frac{r_{ab}}{c}){}^{(a)}\alpha^{\nu_s\nu_v}_{+-}(\tau_s,\tau_v)\bigg]
\end{align}
Where we have defined
\begin{align}
\tilde{E}_{21}^{\nu'_v}(\mathbf{r}_b,t)=\sum_{\nu_1\zeta_1}\sum_{\nu_2\zeta_2}\int d\tau_2d\tau_1E_2^{\nu_2\zeta_2}(\mathbf{r}_b,\tau_2)E_1^{\nu_1\zeta_1}(\mathbf{r}_b,\tau_1){}^{(b)}\beta^{\nu'_v\nu_2\nu_1}_{+--}(t,\tau_2,\tau_1)
\end{align}
Note that we now use 
\begin{align}
{}^{(a)}\alpha_{+-}^{\nu_i\nu_j}(t_i,t_j)=\langle \mathcal{T}V_+^{\nu_i}(t_i)V_-^{\nu_j}(t_j)\rangle_a
\end{align}
which is technically a redundant definition since $\alpha=\bar{\alpha}$ because $Tr[V_-\dots]=0$.  In these expressions, we have used the effective $E$ fields (denoted by $\tilde{E}$) to illustrate how the local-field effects may be thought of as due to the polarization from one molecule serving as an effective field for the other molecule.  This result may also be expressed by defining a total response function:
\begin{align}
S^{(2)}(T_s,T_2,T_1)=\Im\bigg[\frac{4\pi}{\hbar^5}\sum_{ab}\sum_{\nu_i\zeta_i}\int d\tau_sd\tau_2d\tau_1d\tau_vE_s^{\nu_s\dagger}(\tau_s)E_2^{\nu_2\zeta_2}(\tau_2)E_1^{\nu_1\zeta_1}(\tau_1)\mathcal{R}_{ab}^{s21}(\tau_s,\tau_2,\tau_1,\tau_v)\bigg]
\end{align}
We may then simply read off the response function resulting in Eqn.~\ref{eq:RTsecond}.

\end{document}